\documentclass[%
 reprint,
 amsmath,amssymb,
 aps,
]{revtex4-1}

\usepackage{color}
\usepackage{graphicx}
\usepackage{dcolumn}
\usepackage{bm}

\usepackage{dsfont}
\usepackage{dsfont}
\usepackage{float}

\newcommand{\subfigimg}[3][,]{%
  \setbox1=\hbox{\includegraphics[#1]{#3}}
  \leavevmode\rlap{\usebox1}
  \rlap{\hspace*{1pt}\raisebox{\dimexpr\ht1-2\baselineskip}{#2}}
  \phantom{\usebox1}
}

\begin{document}

\preprint{APS/123-QED}

\title{Staircase to Higher-Order Topological Phase Transitions}

\author{P. Cats$^1$, A. Quelle$^1$, O. Viyuela$^{2,3}$, M. A. Martin-Delgado$^4$, C. Morais Smith$^1$}
\affiliation{$^1$Institute for Theoretical Physics, Centre for Extreme Matter and Emergent Phenomena, Utrecht University, Princetonplein 5, 3584CC Utrecht, the Netherlands  \\ 
$^2$Department of Physics, Harvard University, Cambridge, MA 02318, USA\\
$^3$Department of Physics, Massachusetts Institute of Technology, Cambridge, MA 02139, USA\\
$^4$Departamento de F\' isica te\'orica I, Universidad Complutense, 28040 Madrid, Spain.}

\begin{abstract}
We find a series of topological phase transitions of increasing order, beyond the more standard second-order phase transition in a one-dimensional topological superconductor. The jumps in the order of the transitions depend on the range of the pairing interaction, which is parametrized by an algebraic decay with exponent $\alpha$. Remarkably, in the limit $\alpha = 1$ the order of the topological transition becomes infinite. We compute the critical exponents for the series of higher-order transitions in exact form and find that they fulfill the hyperscaling relation. We also study the critical behaviour at the boundary of the system
and discuss potential experimental platforms of magnetic atoms in superconductors.

\end{abstract}

\pacs{Valid PACS appear here}
\maketitle

\noindent {\it I. Introduction.---} Quantum phase transitions (QPTs) are one of the cornerstones in modern condensed matter physics \cite{Vojta_03,SachdevBook}. These are phase transitions where the variation of a physical parameter (coupling constant) drives a transition from one state of matter with a certain order (phase) to another one with different physical properties. These transitions stem from the quantum fluctuations in the energy spectrum of
the system, due to the Heisenberg uncertainty principle. Ideally, they may occur in the absence of thermal fluctuations, at zero temperature.
Famous examples of QPTs comprise the superfluid to Mott-insulator transition \cite{Fischer_et_al_1989,Greiner_et_al02}, the insulator  to superconductor transition in cuprates \cite{Sachdev_03}, metal-insulator transitions in disordered two-dimensional (2D) electron gases \cite{Punnoose_et_al05}, etc. 

Standard QPTs fit into the Landau theory of phase transitions \cite{Landau37} where different phases can be discriminated by the symmetry of an order parameter. 
Remarkably, there also exist non-standard phase transitions in topological systems \cite{Wen90}. 
They go beyond the standard classification of quantum phases, since they can neither be described by a local order parameter nor by the breaking of a symmetry at the phase-transition point. On the contrary, they are characterized by a global order parameter, which is a topological invariant of the system \cite{Kosterlitz_et_al72,Kosterlitz_et_al73,Thouless_et_al82,Thouless_et_al85,Haldane83,Haldane88}. 

A general criterion to classify phase transitions was put forward initially by Ehrenfest, who associated the degree (order) of the phase transition  to the lowest derivative of the free energy that is discontinuous at the transition point \cite{Ehrenfest33}. 
Later on, phase transitions were identified that fell outside the Ehrenfest classification, like the logarithmic singularity in the specific heat of the Onsager solution
to the Ising model in 2D \cite{Onsager44}. 
This led to a simplified binary classification of phase transitions into first-order and continuous phase transitions \cite{Vojta_03,SachdevBook}.
Though the Ehrenfest criterion is not fully general, it can still be adapted \cite{Pippard57,Jaeger98} to define the order of the phase transition when non-analyticities in the free energy 
are encountered. This will be the case for the series of topological
phase transitions found in our Rapid Communication.

Examples of higher-order phase transitions do not abound.
One instance is found in the large-$N$ approximation of lattice
QCDs in 2D, that happens to be of third order \cite{Gross80}.
Another example appears in the exact solution of the 2D Ising model
coupled to quantum gravity where the transition is also third
order \cite{Kazakov86}. Recently, also a phase transition of infinite order was found in a long-range spin mode \cite{Maghrebi_et_al17}.

When it comes to topological phases of matter \cite{BernevigBook}, only a few examples of first- \cite{Imriska_et_al16,Kempkes_et_al16}, second- \cite{Kitaev01,Kempkes_et_al16},  third- \cite{Rombouts_et_al10,Quelle_et_al16,Kempkes_et_al16} and fourth-order \cite{Kempkes_et_al16}  topological phase transitions have been found and, to the best of our knowledge, never higher than that. 
Therefore, the question of whether higher-order topological phase transitions can appear in symmetry-protected topological systems, and of whether the bulk and boundary may behave differently, remains open.

We focus our Rapid Communication on a one-dimensional
(1D) model of topological superconductors exemplified by the Kitaev chain \cite{Kitaev01}. An interesting extension of this model includes hopping and pairing interactions that are long range \cite{Vodola_et_al14}.
 The study of the topological phases of this long-range
Kitaev chain (LRKC) has revealed a very rich structure, including the existence of topological massive Dirac edge states when the pairing is long-range enough \cite{Viyuela_et_al16}.
When the model is 2D, the propagating Majorana modes get enhanced by long-range hopping and pairing \cite{Viyuela_et_al17}. 
This opens new perspectives for their experimental realization.

In this Rapid Communication, we show that the LRKC
displays a staircase of higher-order topological phase transitions
as we vary the long-range decaying exponent $\alpha$ of the pairing interaction. Remarkably, when $\alpha\rightarrow1$ the order of the phase transition becomes infinite. By considering the ground-state energy, we determine the order of the phase transition, the corresponding critical exponents and check that they satisfy the hyperscaling relation.
Moreover, using correlation functions of the bulk and boundary combined with a thermodynamic approach \cite{Quelle_et_al16,Kempkes_et_al16}, we also analyse the critical behaviour at the boundary where a transition from a system with Majorana Zero Modes (MZMs) to non-local massive Dirac fermions occurs. 
Remarkably, in the LRKC the bulk and boundary topological phase transition decouple, and the universality found in Ref.~\cite{Kempkes_et_al16} where the phase transitions in the bulk was always one order higher than at the edges, no longer holds.


\begin{figure*} 
\centering
\begin{tabular}{@{}p{0.45\linewidth}@{\quad}p{0.45\linewidth}@{}}
\centering
\subfigimg[width=1\linewidth]{(a)}{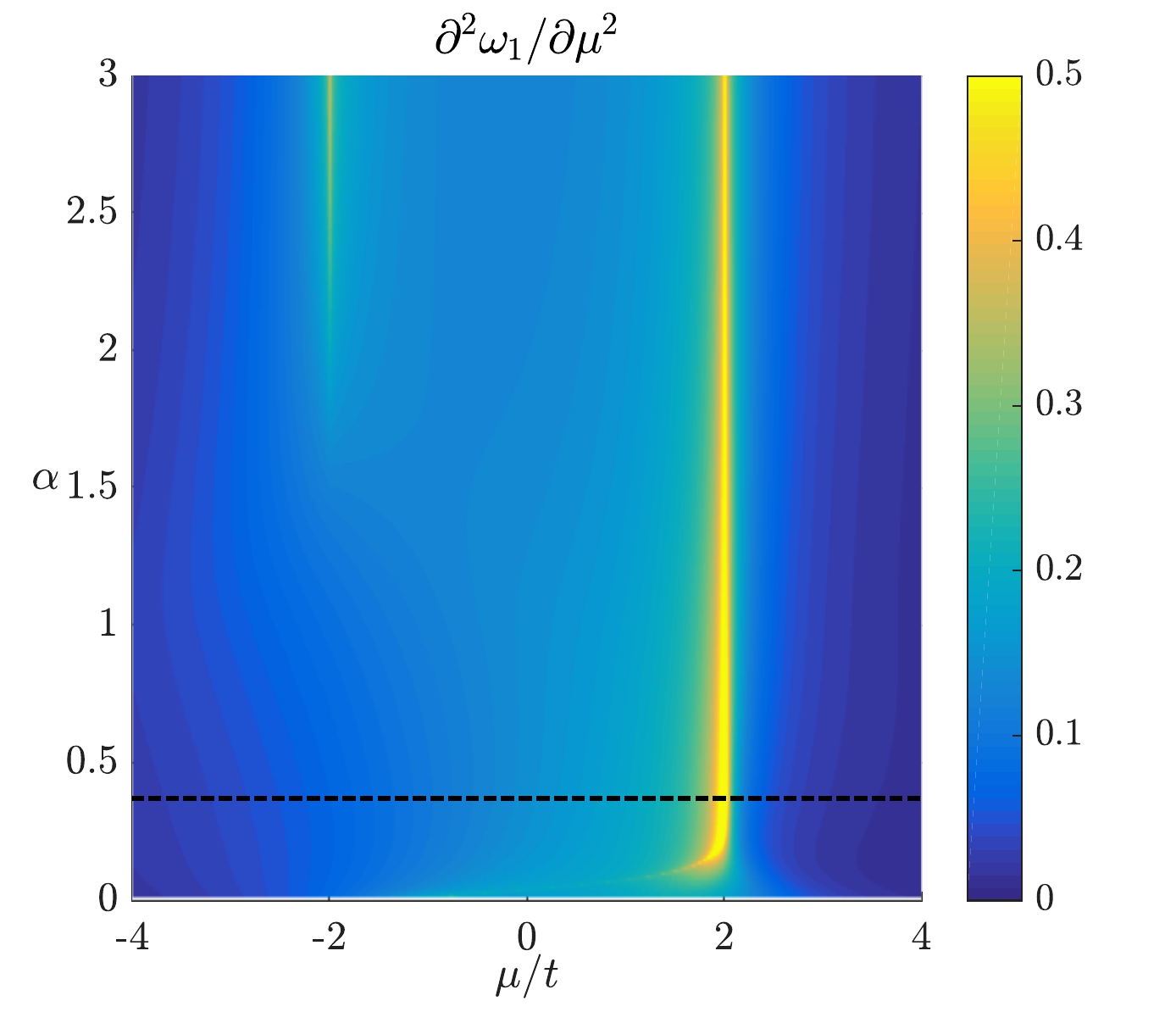} &
\subfigimg[width=1\linewidth]{(b)}{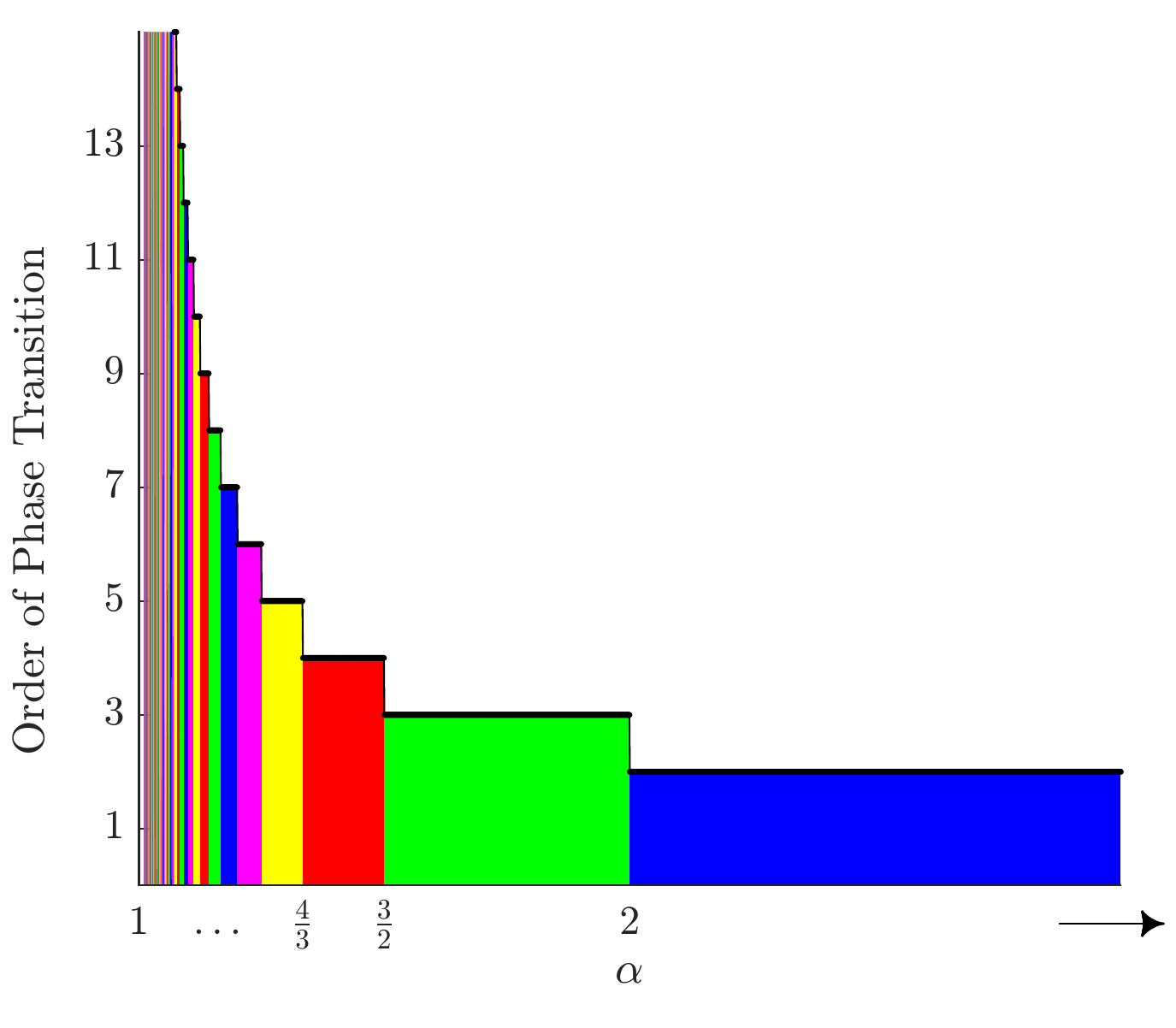} 
\end{tabular}
\caption{(a) Second derivative of the bulk thermodynamic potential, obtained from Eq.~\eqref{Eq:dw1dmu} for the LRKC in the $\mu/t$-$\alpha$ plane. The dashed line indicates the value of $\alpha$ below which our numerical results are not accurate anymore. (b) Staircase of higher-order topological phase transition at $\mu/t = -2$, indicating how the order of the transition increases as one approaches $\alpha = 1$. 
\label{Fig:LR_Kit_surf_dw1dmu}}
\end{figure*}

\noindent {\it II. The model.---} The Hamiltonian of the LRKC \cite{Vodola_et_al14,Viyuela_et_al16} with $N$ sites  reads 
\begin{align}\label{Eq:Ham}
\hat{H}=&-\mu\sum_{j=1}^N\left(c_j^\dagger c_j-\frac{1}{2}\right)-t\sum_{j=1}^{N-1} \left( c_{j}^\dagger c_{j+1} +c_{j+1}^\dagger c_j\right) +\nonumber\\ &+\Delta \sum_{j\neq l} \frac{1}{|j-l|^\alpha} \left( c_j c_l + c_l^\dagger c_j^\dagger\right),
\end{align}
where $c_j$ ($c_j^\dagger$) is the fermionic annihilation (creation) operator for site $j$, $\mu$ the chemical potential, $t$ the hopping parameter, $\Delta$ the pairing amplitude and $\alpha$ the parameter characterising the range of the interaction. Long-range hopping terms can be also considered, but they do not provide novel topological phases \cite{Viyuela_et_al16}. Thus, we may consider purely short-range hopping without loss of generality. The spectrum of the LRKC with periodic boundary conditions is given by \cite{Vodola_et_al14,Viyuela_et_al16}
\begin{align}\label{Eq:spectrum}
&E_k=\pm \sqrt{\epsilon_k^2+4\Delta^2f_\alpha^2(k)}, &f_\alpha(k)=\sum_{l=1}^{N-1}\frac{\sin(kl)}{|l|^\alpha},
\end{align}
where $\epsilon_k=-\mu-2t\cos(k)$.
For $\alpha\rightarrow \infty$, this model has a well-defined limit to the  short-range Kitaev chain (SRKC) \cite{Kitaev01}, which is known to display a topological phase for $|\mu/t|<2$, characterized by the presence of Majorana zero modes (MZMs) at the edges. For $|\mu/t|>2$, a trivial phase is found instead. 

The LRKC exhibits even more exotic behaviour than its short-range counterpart \cite{Vodola_et_al14,Viyuela_et_al16,Luca_16}. The long-range terms give rise to the function $f_\alpha(k)$ defined in Eq.~\eqref{Eq:spectrum}, which is discontinuous at $k=0$ for $\alpha<1$, whereas its derivative is discontinuous for $\alpha<2$.
Therefore, the physics of the model drastically depends on $\alpha$.

For $\alpha>2$, the LRKC behaves similarly to the SRKC, i.e. there are MZMs in the topological phase. 
For $\alpha<1$, the physics in the topological phase changes drastically, in that the two Majorana modes at the edge {merge into a non-local Dirac fermion that acquires mass} provided that $\mu/t<2$.
Finally, for $1<\alpha<2$, the topological phase diagram becomes more intricate, and the winding number becomes ill-defined, as discussed in Ref. \cite{Viyuela_et_al16}. The critical behaviour at $\mu/t=-2$ in the bulk changes with $\alpha$, whereas the one at $\mu/t=2$ remains the same. Contrarily to the bulk, the boundary of the LRKC behaves still in the same way as the SRKC for $3/2<\alpha<2$. However, for $\alpha<3/2$ one finds, in addition to the MZMs for $|\mu/t|<2$, non-local massive edge states for $\mu/t<-2$ \cite{Viyuela_et_al16,Sup.Mat.}. A disorder analysis of the sector $1<\alpha<2$ (see "The  edge states for $\alpha<3/2$" in the Supplemental Material \cite{Sup.Mat.}) shows the robustness of these massive edge states to static disorder.

In order to understand the nature of the topological phase transitions at $\mu/t=\pm2$ within the different topological sectors, we investigate their thermodynamic properties using correlation functions. As it turns out, the order of the phase transition at $\mu/t=2$ does not change with $\alpha$, but we find extraordinary behaviour for the order of the phase transition at $\mu/t=-2$, in the form of a staircase of higher-order topological phase transitions towards $\alpha\rightarrow1$. Before we show these results, let us first introduce our method.

\noindent {\it III. Thermodynamic analysis.---} To classify the phase transitions of the LRKC, we use an adapted Ehrenfest classification \cite{Ehrenfest33,Pippard57,Jaeger98}, in which one considers the grand potential $\Omega$ and assigns the order of the phase transition according to the derivative for which the grand potential has a divergence or a discontinuity. The  grand potential can subsequently be decomposed into a bulk term $N\omega_1$, which scales linearly with the system size, and a residual term $\omega_0$, which contains the finite-size and boundary effects, i.e. $\Omega=N\omega_1+\omega_0$. To obtain these contributions, we consider the derivative of the grand potential $\Omega$ w.r.t. $\mu$, such that we can relate it directly to the correlation functions
\begin{align}\label{Eq:Omega}
\frac{\partial \Omega}{\partial \mu}=\frac{\partial }{\partial \mu}\left( -\frac{1}{\beta} \ln \mathrm{Tr}\left[ e^{-\beta \hat{H}}\right]\right)=\left\langle \frac{\partial \hat{H}}{\partial \mu}\right\rangle,
\end{align}
where $\langle \hat{A} \rangle:={\mathrm{Tr}[ \hat{A} e^{-\beta \hat{H}}]}/{\mathrm{Tr}[ e^{-\beta \hat{H}}]}$.
This thermodynamic analysis is especially well-suited for symmetry-protected topological systems both at zero \cite{Quelle_et_al16} and finite temperatures \cite{Kempkes_et_al16,Viyuela_et_al14}.
To explicitly find $\omega_1$ and $\omega_0$, we consider an infinitely long and periodic chain at zero temperature with grand-potential density 
\begin{align*}
\omega=\frac{\Omega_p}{N}=-\int_{-\pi}^{\pi} \mathrm{d}k E_k,
\end{align*}
where $p$ stands for periodic. This integral is bounded because the spectrum is finite for $\alpha>1$, and diverges at most as $1/k^{1-\alpha}$ for $\alpha<1$. Hence, $\omega$ is finite for all $\alpha$ and does not depend on the system size in this limit. Similarly, the on-site correlation function $\langle c_r^\dagger c_r\rangle$ does not depend on the system size when $N$ is large enough.  Thus, we may add and subtract $\partial \Omega_p/\partial \mu$ to Eq.~\eqref{Eq:Omega}, to find
\begin{align}\label{Eq:dO_dOp}
\frac{\partial \Omega}{\partial \mu}&=\frac{\partial \Omega_p}{\partial \mu}+\left(\frac{\partial \Omega}{\partial \mu}-\frac{\partial \Omega_p}{\partial \mu}\right).
\end{align}
Using Eqs.~\eqref{Eq:Ham},~\eqref{Eq:Omega}, and~\eqref{Eq:dO_dOp}, we then obtain
\begin{align*}
\frac{\partial \Omega}{\partial \mu}&=N\langle c_r^\dagger c_r\rangle+\sum_j \left( \langle c_j^\dagger c_j\rangle - \langle c_r^\dagger c_r \rangle\right),
\end{align*}
where $\langle c_r^\dagger c_r\rangle$ are the on-site correlation functions for the infinitely long periodic chain 
and $\langle c_j^\dagger c_j \rangle$ are calculated 
for the finite chain. 
Hence, we can read off
\begin{align}
\frac{\partial \omega_1}{\partial \mu}&= \langle c_r^\dagger c_r\rangle, \label{Eq:dw1dmu}\\
\frac{\partial \omega_0}{\partial\mu}&=\sum_j \left( \langle c_j^\dagger c_j\rangle - \langle c_r^\dagger c_r \rangle\right):= \sum_j \Lambda(j). \label{Eq:dw0dmu}
\end{align}
The extensive bulk term $\partial\omega_1/\partial \mu$ is simply given by the on-site correlation functions for the infinitely long periodic chain and the residual term $\partial \omega_0/\partial \mu$ is the sum of the difference $\Lambda(j)$ between the on-site correlation functions in the periodic and in the  finite chain. As a consequence, the residual contribution $\omega_0$ contains all the subleading terms in $N$, and therefore includes $\log(N)$, constant, and $1/N$ terms, to name just a few.
For large system sizes, one can only consider the logarithmic and the constant term and neglect all other subleading contributions.

For the SRKC, it suffices to only consider the constant term
\cite{Quelle_et_al16,Kempkes_et_al16}. This leads to a first-order phase transition at the boundaries \citep{Kempkes_et_al16}, which is due to the appearance/disappearance of the Majorana edge states. The second-order phase transition in the bulk is due to a gap closing at the critical points $\mu/t=\pm 2$. Let us now focus on the LRKC where higher-order topological phase transitions will arise.

\noindent {\it IV. Higher-order bulk phase transitions.---} 
We analyse the zero-temperature behaviour of the bulk grand-potential density defined in Eq. (5) as a function of the long-range exponent $\alpha$ and the chemical potential $\mu$. Although we concentrate here on the zero-temperature behaviour, the method itself is generic and could be applied to finite temperatures. The results are shown in Fig.~\ref{Fig:LR_Kit_surf_dw1dmu}. There is a second-order phase transition at $\mu/t = 2$ separating the topological and trivial phases for every value of $\alpha$ [see Fig.~\ref{Fig:LR_Kit_surf_dw1dmu}(a)], precisely as it is the case for the SRKC. Note the behaviour below the dashed line around $\alpha\approx 0.3$ near $\mu/t=2$, where the transition line makes a turn and does not go all the way towards $\alpha=0$. This is merely an artifact due to numerical limitations, since the correlations become too long-ranged, and one needs very large system sizes to suppress this effect.

On the other hand, for $\mu/t = -2$ the behaviour of the phase transition changes drastically, depending on the value of $\alpha$, and further analytical calculations are needed. 
Since the non-analytical behaviour of the bulk term in the grand-potential density $\omega_1$ is given by the $k=0$ mode, we make the separation $\omega_1=F+G$, where $F$ is the integral around $k=0$, containing all the non-analyticities, and $G$ is the integral over the remaining part of the Brillouin zone. In this way, we can consider only $F$ to describe the non-analytical part of $\omega_1$, i.e. the information about the order of the phase transition. To calculate $F$, one can expand the spectrum $E_k$ in Eq.~\eqref{Eq:spectrum} around $k = 0$ and integrate it for $k\in (0, \varepsilon)$, where $\varepsilon$ is sufficiently small for the expansion in $k$ to be valid. From this expansion, we can also extract the critical exponent $\tilde{\alpha}$ defined by 
$ \Omega \propto m^{2-\tilde{\alpha}}$, the dynamical exponent $z$  defined by $E_{k}(m=0)\propto k^{z}$ and the critical exponent $\nu$ defined by $E_{k=0}(m)\propto m^{z\nu}$ \cite{SachdevBook}, where $m=\mu/t+2$ denotes the distance from the critical point.
The leading term for $\alpha>2$ (valid also for the SRKC) casts the form
\begin{align*}
F(m,\alpha>2):=\int_0^\varepsilon \mathrm{d}k \sqrt{m^2+k^2}\propto m^2\log |m|.
\end{align*}
This function is divergent in its second derivative at $m=0$ for all $\alpha>2$, hence we find a second-order phase transition. For $1<\alpha<2$, the leading term is given by
\begin{align}\label{Eq:F_exp_alpha_LEQ_2}
F(m,1<\alpha&<2)\approx \int_0^\varepsilon \mathrm{d}k \sqrt{m^2+k^{2(\alpha-1)}}\\
&\propto \Gamma\left(-\frac{\alpha}{2(\alpha-1)}\right)\Gamma\left(\frac{2\alpha-1}{2(\alpha-1)}\right)|m|^{\frac{\alpha}{\alpha-1}},\nonumber
\end{align}
where the last line is not defined if one of the $\Gamma$ functions diverge, which happens when $\alpha=n/(n-1)$ where $n\in \mathds{N}$. For $\alpha=2n/(2n-1)$, one finds, instead of Eq.~\eqref{Eq:F_exp_alpha_LEQ_2}, the relation $F\propto m^n \log(|m|)$ , which is divergent in its $n$-th derivative at $m=0$. For $\alpha=(2n-1)/(2n-2)$, there will be a discontinuity in the $2n-1$-th derivative at $m=0$, because in that case $F\propto |m|^{2n-1}$. Hence, for any $\alpha=n/(n-1)$, we find a $n$-th order phase transition. If $\alpha$ is in between these values, then the power of $|m|$ in Eq.~\eqref{Eq:F_exp_alpha_LEQ_2} is a non-integer value, meaning that one can differentiate it until its power is negative and it becomes divergent at $|m|=0$. For example, if $3/2<\alpha<2/1$ the exponent of $|m|$ lies between 2 and 3, which means that the third derivative is divergent at $m=0$. For $4/3<\alpha<3/2$, the exponent of $|m|$ lies between 3 and 4, which means that the fourth derivative is divergent, and so forth. 

Therefore, we find a staircase behaviour such that the order of the topological phase transition increases step-wise at the points $\alpha=n/(n-1)$ upon lowering $\alpha$ from  $\alpha=2$ to $\alpha=1$ [see Fig.~\ref{Fig:LR_Kit_surf_dw1dmu}(b)]. This implies that by tuning the exponent $\alpha$ we may address qualitatively different thermodynamic behaviour by inducing higher-order topological phase transitions.

In addition, the critical exponents that follow from this analysis read $\tilde{\alpha}=(\alpha-2)/(\alpha-1)$, $z=\alpha-1$ and $\nu=1/(\alpha-1)$. This is consistent with the hyperscaling relation $2-\tilde{\alpha}=\nu(d+z)$, where $d$ denotes the dimensionality of the model ($d=1$ in our case) \cite{BarberBook,Mucio}.

A remarkable effect occurs in the limit $\alpha=1$, as the topological phase transition becomes of infinite order. 
Instead, for $\alpha<1$, there is no longer a phase transition at $\mu/t=-2$ because the long-range pairing causes the whole chain to be correlated, thus gapping the edge mode everywhere. This can also be clarified from Eq.~\eqref{Eq:F_exp_alpha_LEQ_2}, where the integrand can be expanded as $\sqrt{m^2+k^{2(\alpha-1)}}= k^{-(1-\alpha)}+k^{1-\alpha}m^2/2+ \mathcal{O}(k^{1-\alpha}(k^{2(1-\alpha)}m^2)^2)$, such that $F(m,\alpha<1)$ exhibits no non-analyticities. Thus, $\alpha=1$ constitute a critical-end point in the $\alpha-\mu/t$ quantum phase diagram. This behaviour is consistent with the Dirac Sector found in Ref.~\cite{Viyuela_et_al16}.

\begin{figure}[t]
\centering
\includegraphics[width=\columnwidth]{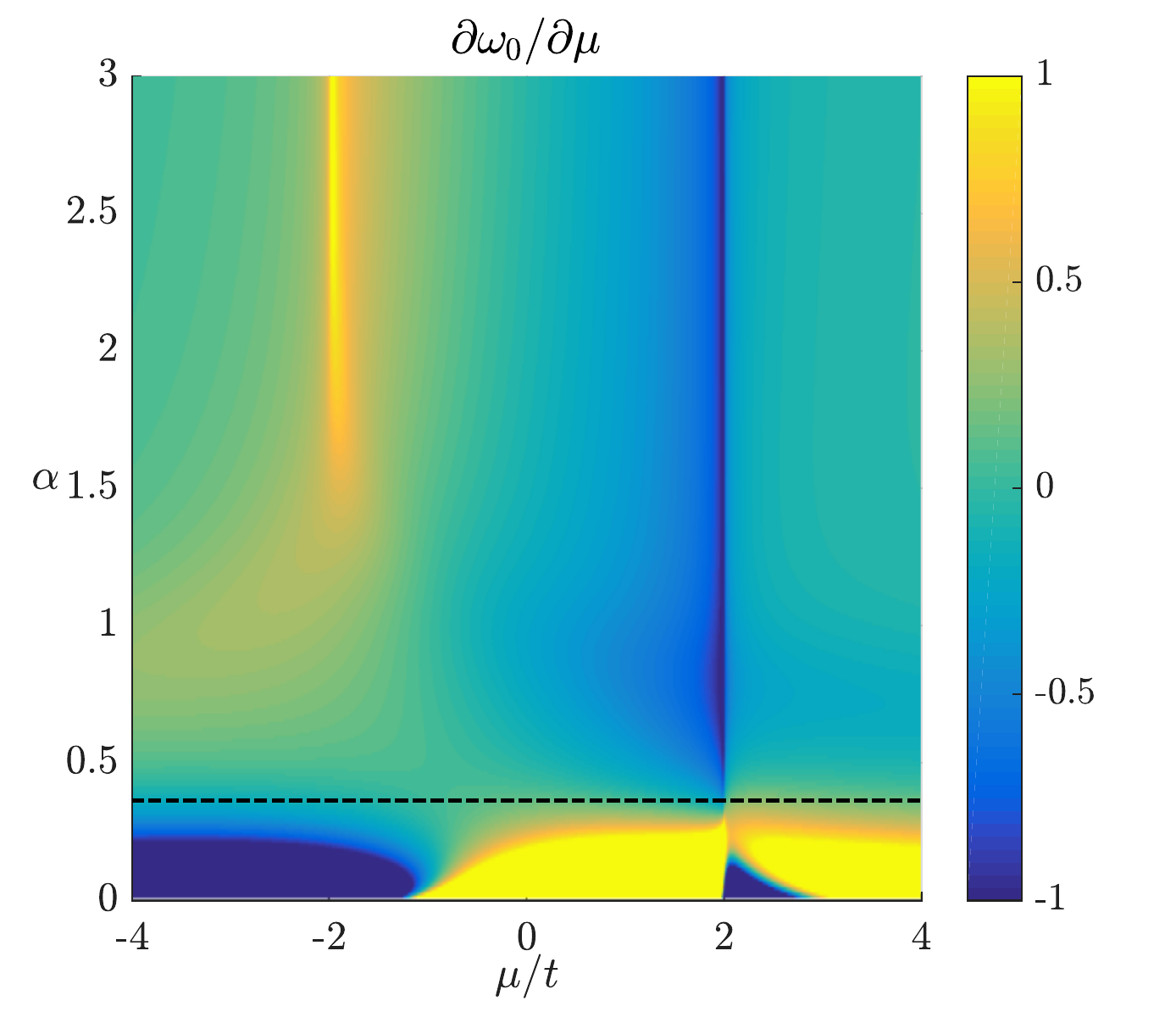} 
\caption{First derivative of the boundary thermodynamic potential as a function of $\mu/t$, given by Eq.~\eqref{Eq:dw0dmu}. The dashed line indicates the value of $\alpha$ below which our numerical results are not accurate anymore. Although the topological phase transition at $\mu/t =  2$ remains of first order in the entire regime of $\alpha$, in the thermodynamic limit the phase transition at $\mu/t = -2$ terminates at a quantum critical point located at $\alpha = 1$. However, numerical limitations prevent us from going below $\alpha = 3/2$. 
\label{Fig:LR_Kit_surf_dw0dmu}}
\end{figure}

\noindent {\it V. Boundary phase transition.---} The separation between the bulk and residual contribution to the grand potential allows us to investigate the behaviour of both independently. Using Eq.~\eqref{Eq:dw0dmu}, we calculate numerically $\partial \omega_0/\partial \mu$ in the $\alpha-\mu/t$ plane and find the result given in Fig.~\ref{Fig:LR_Kit_surf_dw0dmu} (a) for $N=200$. Below the dashed line, the results are not accurate due to numerical limitations, as was the case for the bulk. Along the line $\mu/t=2$, there is a clear indication of a first-order phase transition for all values of $\alpha$ (dark-blue line). However, for $\mu/t=-2$ there is only a clear indication of a first-order phase transition down to $\alpha=3/2$ (bright-yellow line), below which the boundaries of the phase transition blur out. The reason for this is that for short-range models, when the system is large enough (although finite), the features characterizing the phase transition are so sharp, that one can confidently draw conclusions that would - strictly speaking - only hold for infinite systems. However, when the model becomes long-range, this is no longer the case (see the end-to-end correlations analysis in the Supplemental Material \cite{Sup.Mat.}).



\noindent {\it VI. Conclusions and outlook.---} We discovered a staircase of higher-order topological phase transitions in a long-range Kitaev chain. We have shown that the order of the topological phase transition increases step-wise with the long-range decaying exponent $\alpha$ of the pairing interaction. In the limit $\alpha=1$, we remarkably found an infinite-order phase transition.
By separating the bulk from the residual contribution in the grand potential and performing a thermodynamic analysis, we have established not only the order of the topological phase transitions, but also the corresponding critical exponents and checked that they satisfy the hyperscaling relation. Moreover, we have also studied the critical behaviour at the boundary, where there is a transition from a topological phase with MZMs to another topological phase with non-local massive Dirac edge modes \cite{Viyuela_et_al16}. 

For the long-range Kitaev chain, the correlation functions decay algebraically at long distances, and exponentially at short distances \cite{Vodola_et_al14,Luca_16}. Hence, the system is critical, and 
the  correlation length can no longer be straightforwardly defined \cite{Luca_16}. Although the algebraic
term in the correlation functions (which gives the quasi long-range order) is present for all $\alpha$, it becomes important around the region where the winding number becomes
ill-defined. Therefore, both the criticality and the
ill-defined winding number arise due to the relevance of
long-range effects at small $\alpha$. We would like to emphasise that our results do not depend on the definition of the correlation length in any way, nor on the correlation
length itself, in contrast to scaling theories, where the
scaling of the correlation length is used to determine the
critical exponents. This is one of the main advantages of
our approach, which allows us to describe even critical
systems.

We determine the critical exponents by analysing the behaviour of the grand-potential density at the critical point, and  characterise thus the topological phase transition. Although the results are applied here at zero temperature, the formalism is generic and may also be used at finite temperatures \cite{Kempkes_et_al16}. 
In this case, one should be able to connect the central charge $c$ obtained from the entanglement entropy \cite{Vodola_et_al14} to the central charge found from the heat capacity $C_V$ at very small temperatures, since $C_V\propto c T$ within a first order expansion in $T$ \cite{Fradkin}. This would allow for an independent verification of the anomalous behaviour of the central charge at $\mu/t=-2$ predicted in Ref. \cite{Vodola_et_al14}.

Spin and fermionic topological systems with long-range interactions have recently attracted much attention \cite{Maghrebi_et_al17,Vodola_et_al14,Viyuela_et_al16,Viyuela_et_al17,Luca_16,Vodola_et_al16,Niui_12,DeGottardi_13,Gong2015_1,Gong2015_2,Tudela_15,Vodola_et_al17,Dutta_17,Pachos_17,Alecce_17,Lepori_et_al_17}. In particular, our long-range model is motivated by current experiments of 1D arrays of magnetic atoms deposited on top of conventional $s$-wave superconductor substrates \cite{Nadj_et_al14,Pascual_et_al16,Ruby_et_al17}.  These arrays of magnetic impurities form subgap states known as Shiba states \cite{Yu65,Shiba68,Rusinov69}. The particular wavefunction of these states have power-law tails that lead to hopping and pairing amplitudes that decay algebraically with the distance. For 3D superconducting substrates (for instance lead as in Ref.~\cite{Nadj_et_al14} and~\cite{Ruby_et_al17}) the decay goes as $1/r$, whereas for 2D substrates (for example 2D transition metals dichalcogenides) the decay goes as $1/\sqrt{r}$. This long-range behaviour of Shiba impurities has already been observed in recent experiments \cite{Menard_et_al15}. Apart from the power-law decay, there is an exponential factor that depends on the coherence length of the superconductor. However, when the length of the chain is small with respect to this coherence length \cite{Pientka2013}, the dominant decay is algebraic, and $p$-wave Hamiltonians with intrinsic long-range pairing are induced \cite{Pientka2013,Nadj_et_al13,Pientka2014,Li_et_al_16,Kaladzhyan_et_al16}. A possible way to tune the decaying exponents, such that the hopping and pairing amplitudes
decay differently could be achieved through Floquet driving fields as proposed in Ref.~\cite{Benito_et_al14}. Thus, the
staircase of higher-order topological phase transitions, found in
our Rapid Communication, could be experimentally detected.

\begin{acknowledgments}
The work by A.Q. and C.M.S. is part of the D-ITP consortium, a program of the Netherlands  Organisation  for  Scientific  Research  (NWO)  that  is
funded by the Dutch Ministry of Education, Culture and Science (OCW).
M.A.M.-D. and O.V. acknowledge financial support from the Spanish MINECO grants FIS2012-33152, FIS2015-67411, and the CAM research consortium QUITEMAD+, Grant No. S2013/ICE-2801. The research of M.A.M.-D. has been supported in part by the U.S. Army Research Office through Grant No. W911N F-14-1-0103.
O.V. thanks Fundaci\'on Ram\'on Areces, Fundaci\'on Rafael del Pino and RCC Harvard. 

\end{acknowledgments}

\clearpage

\section*{\Large Supplemental Material}

\renewcommand{\thefigure}{S\arabic{figure}}
\setcounter{figure}{0}   
\section{The Spectrum}
\begin{figure}[!h]
\centering
\includegraphics[width=0.48\textwidth]{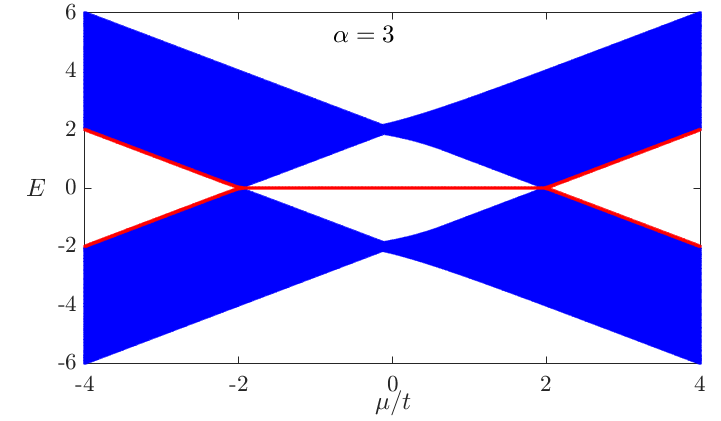}
\includegraphics[width=0.48\textwidth]{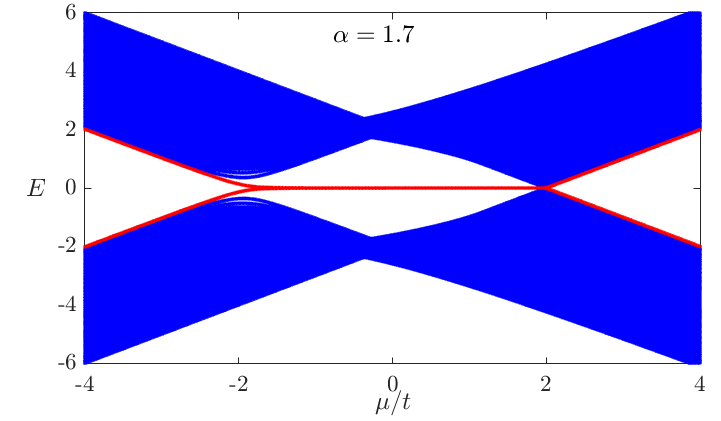}
\includegraphics[width=0.48\textwidth]{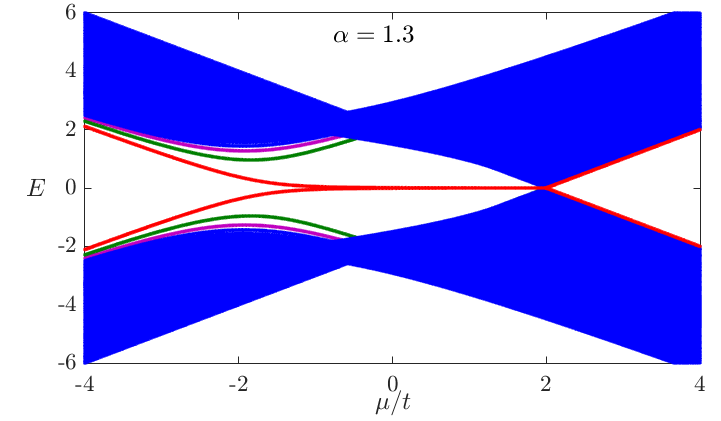}
\includegraphics[width=0.48\textwidth]{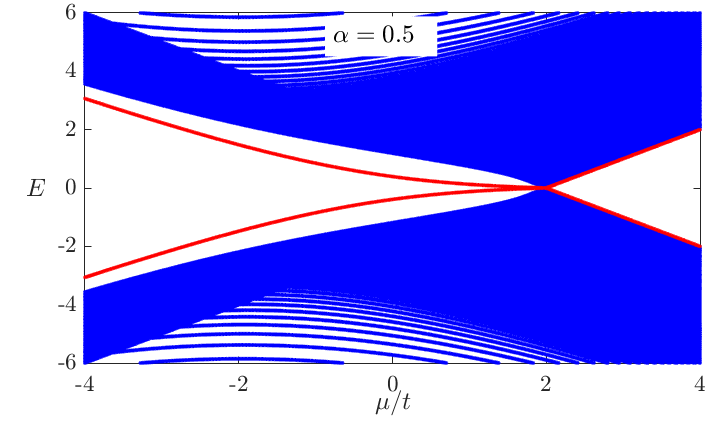}
\caption{The spectrum of the finite chain with $N=200$ for several values of $\alpha$. The lowest energy level is plotted in red. Note that in the plot for $\alpha=1.3$, the first and second excited states that detach from the bulk are coloured green and magenta. \label{Fig:Supp_spectrum}}
\end{figure}

The spectrum of the finite chain with $N=200$ sites for several values of $\alpha$ is given in Fig. \ref{Fig:Supp_spectrum}. A gap starts to open at $\mu/t=-2$ when lowering $\alpha$. In the thermodynamic limit, it will open at $\alpha=1$, but due to the finite system size, the edge states start to hybridize already at slightly higher values of $\alpha$ because of the long-range pairing.  One has to consider enormously large system sizes to see that the bands should not be opening, although the density of states (DOS) approaches zero.
This can be understood by investigating the dispersion and the DOS at $m=0$ and $1<\alpha<2$, for which the former is given by $E_k\approx \sqrt{m^2+k^{2(\alpha-1)}}$. One observes that it vanishes at $m=0$ and $k=0$, i.e. the bandgap closes. The DOS $D(E)$ at that point reads
\begin{align*}
D(E)\propto \frac{\mathrm{d}k}{\mathrm{d}E}\approx \left( \frac{\mathrm{d} k^{\alpha-1}}{\mathrm{d}k} \right)^{-1}\propto k^{2-\alpha}\approx E^{\frac{2-\alpha}{\alpha-1}}.
\end{align*}
Since the power of $E$ is positive for $1<\alpha<2$, we find that $D(E=0)=0$. When $\alpha>2$, the spectrum has a linear dispersion near the phase transition; the DOS becomes energy independent and acquires a finite value.

\section{The  Edge States for $\alpha<3/2$}
\begin{figure*}[t]
\begin{tabular}{@{}p{0.48\linewidth}@{\quad}p{0.48\linewidth}@{}}
\centering
\subfigimg[width=1\linewidth]{(a)}{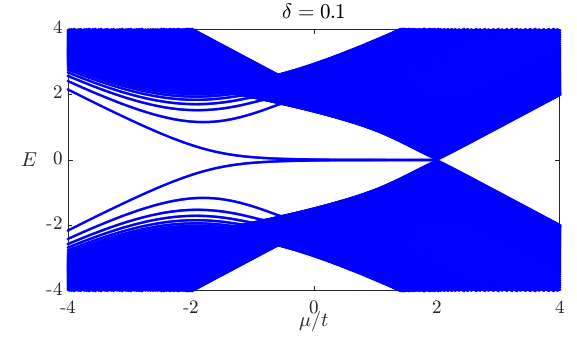} &
\subfigimg[width=1\linewidth]{(b)}{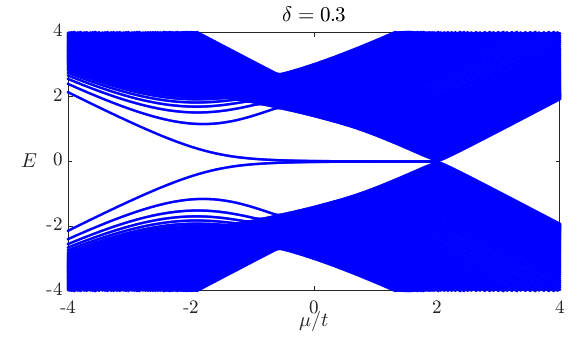}
\end{tabular}
\caption{Energy spectrum for $\alpha=1.3$, $N=100$ sites, averaged over 60 disorder realizations. a) $\delta=0.1$ and b) $\delta=0.3$.}
\label{Fig:spec_dis}
\end{figure*}

\begin{figure*}[t]
\begin{tabular}{@{}p{0.48\linewidth}@{\quad}p{0.48\linewidth}@{}}
\centering
\subfigimg[width=1\linewidth]{(a)}{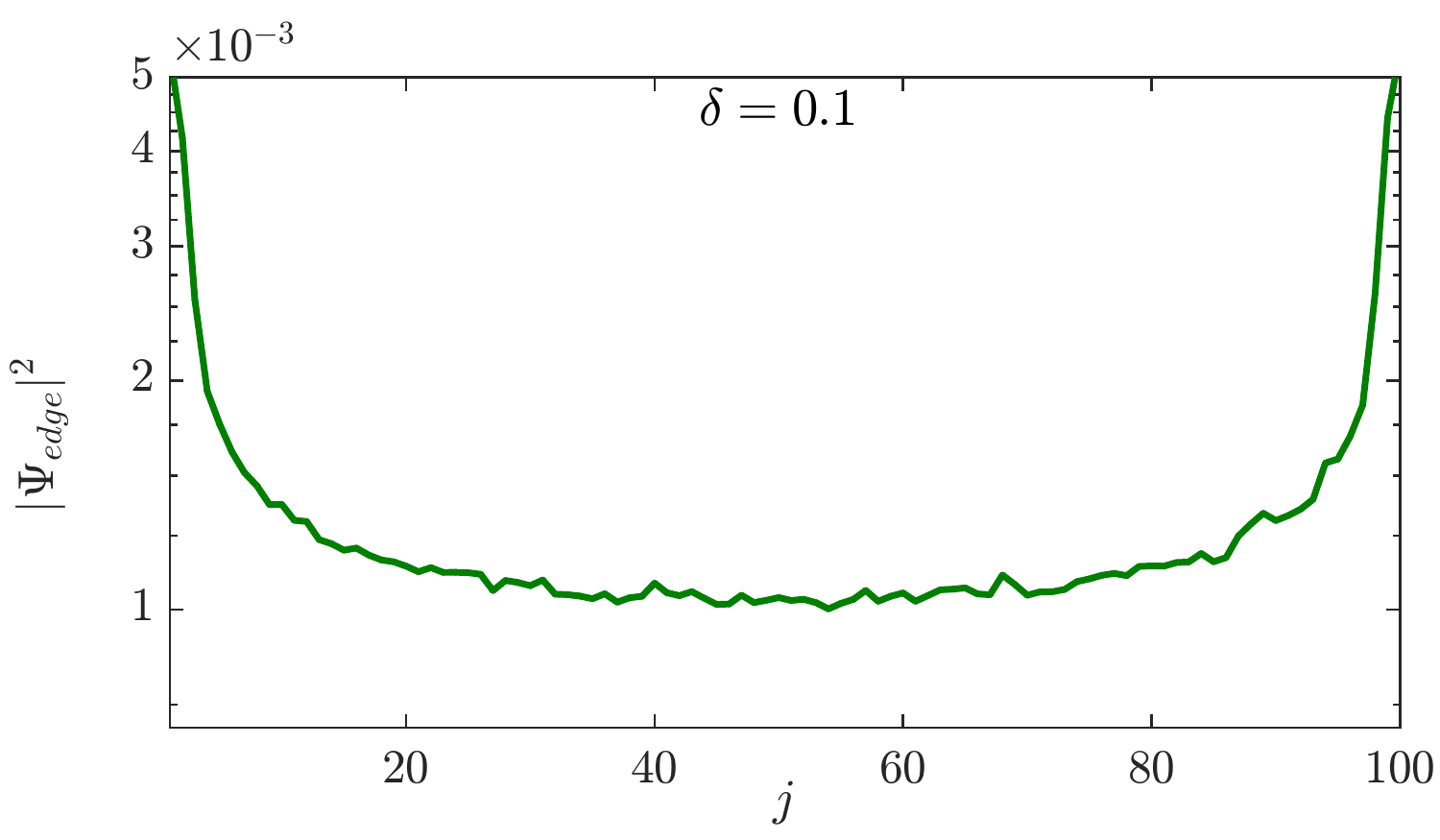} & 
\subfigimg[width=1\linewidth]{(b)}{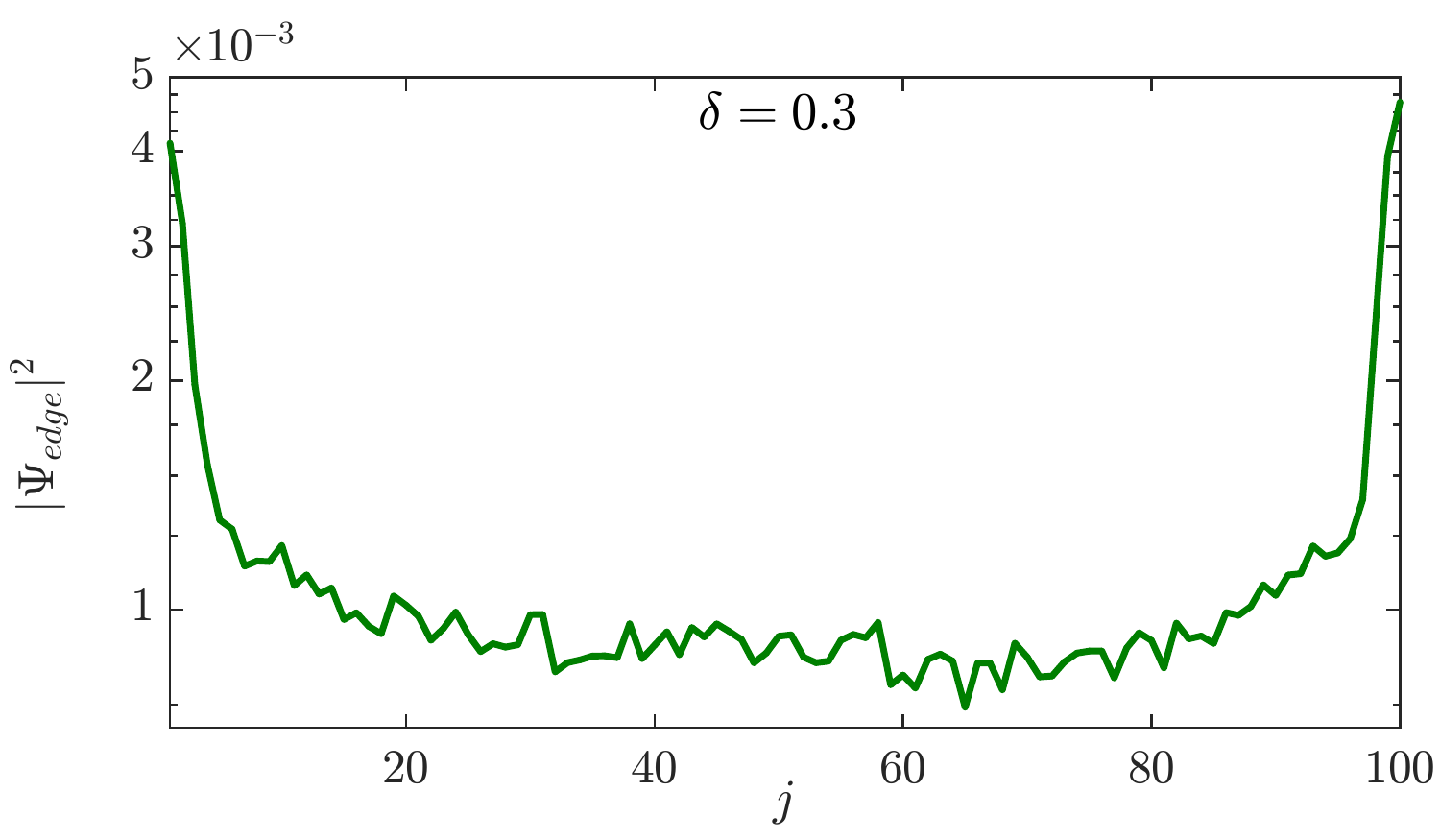}
\end{tabular}
\caption{Edge state probability distribution for $\alpha=1.3$, $\mu/t=-3.0$, $N=100$ sites, averaged over 60 disorder realizations. a) $\delta=0.1$ and b) $\delta=0.3$.}
\label{Fig:edge_dis}
\end{figure*}

In Ref.~\cite{Viyuela_et_al16}, the robustness of the  edge states to disorder is investigated. The disorder term 
\begin{align*}
\hat{H}=\sum_j \epsilon_j c_j^\dagger c_j,
\end{align*}
is added to the Hamiltonian in Eq. (1) in the main text,
where $\epsilon_j$ takes values from a random distribution of zero mean value and width $\delta$. The results, presented in Figs.~\ref{Fig:spec_dis} and \ref{Fig:edge_dis}, indicate that the non-local Dirac fermions with finite energy, are also robust against disorder, i.e. the edge states remain localized at the edges and do not merge into the bulk. Notice that this occurs also outside the topological phase (see Fig. \ref{Fig:spec_dis} and \ref{Fig:edge_dis}, where the results are shown for $\mu/t=-3$). In addition, the energy of the edge states is still separated from the other bulk states, even after considering strong static disorder.  The winding number defined in Ref. \cite{Viyuela_et_al16} becomes ill defined for $\alpha<2$. For $\alpha\in(2,3/2)$, the $y$-component of the winding vector becomes singular at $k=0$, whereas for $\alpha<3/2$ the two components of the winding vector become singular at $k=0$. Interestingly, for $\alpha<1$ it is still possible to define a quantised half-integer winding number that highlights the presence of non-local massive Dirac fermions. For a more elaborated discussion, we refer the reader to Ref. \cite{Viyuela_et_al16}.

\section{The Bulk and Residual contribution to the grand potential} 
\begin{figure*} 
\begin{tabular}{@{}p{0.48\linewidth}@{\quad}p{0.48\linewidth}@{}}
\centering
\subfigimg[width=1\linewidth]{(a)}{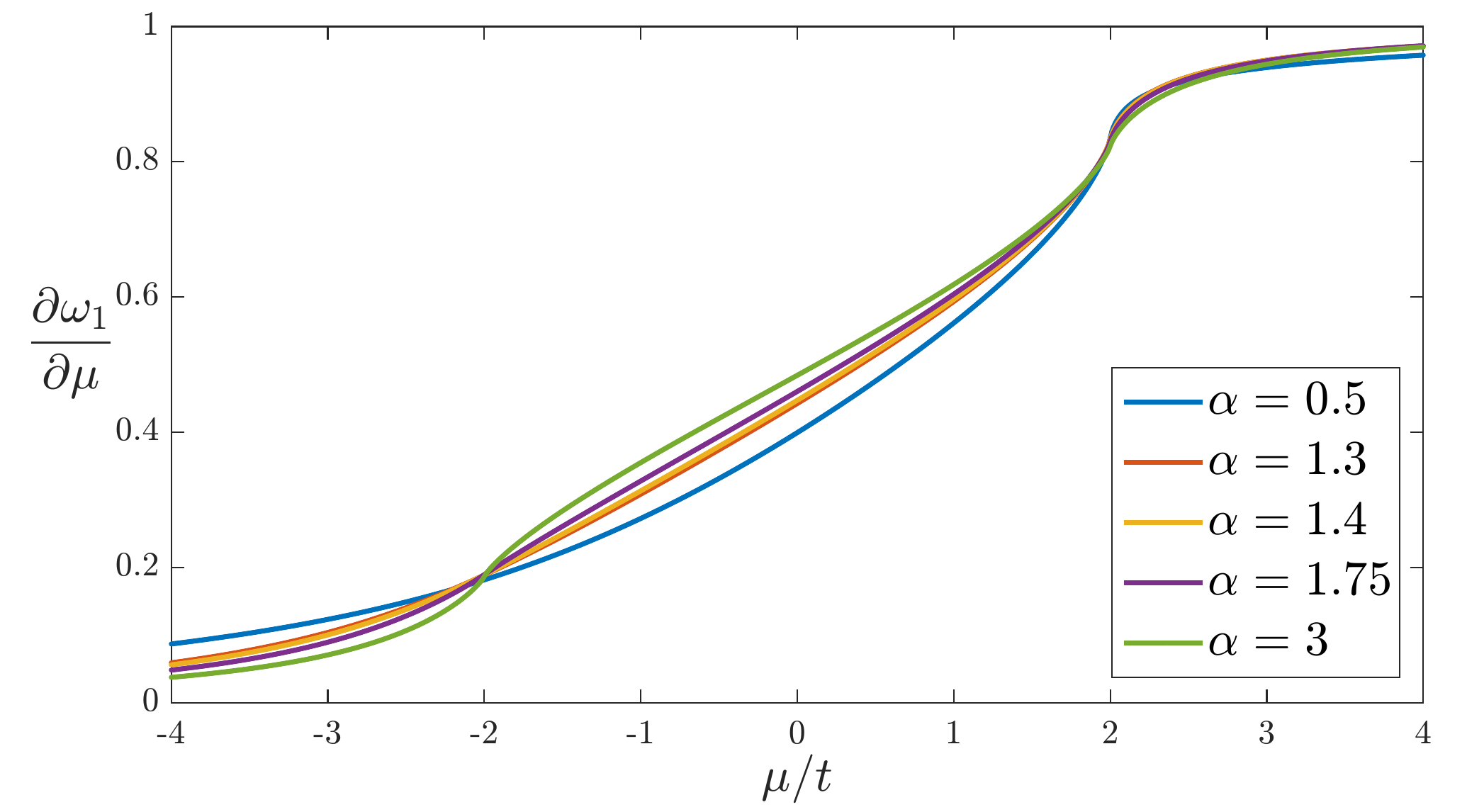} &
\subfigimg[width=1\linewidth]{(b)}{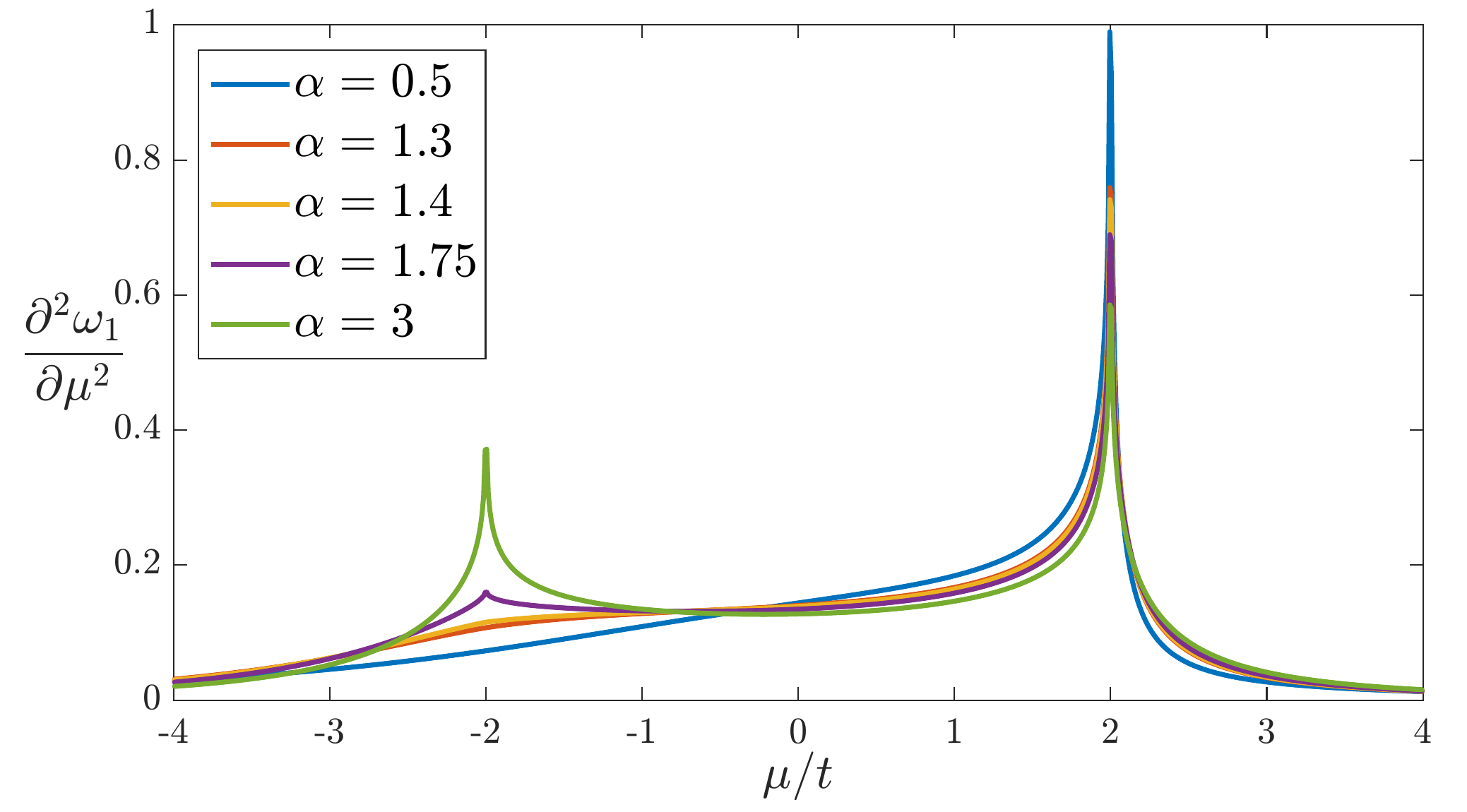}\\
\subfigimg[width=1\linewidth]{(c)}{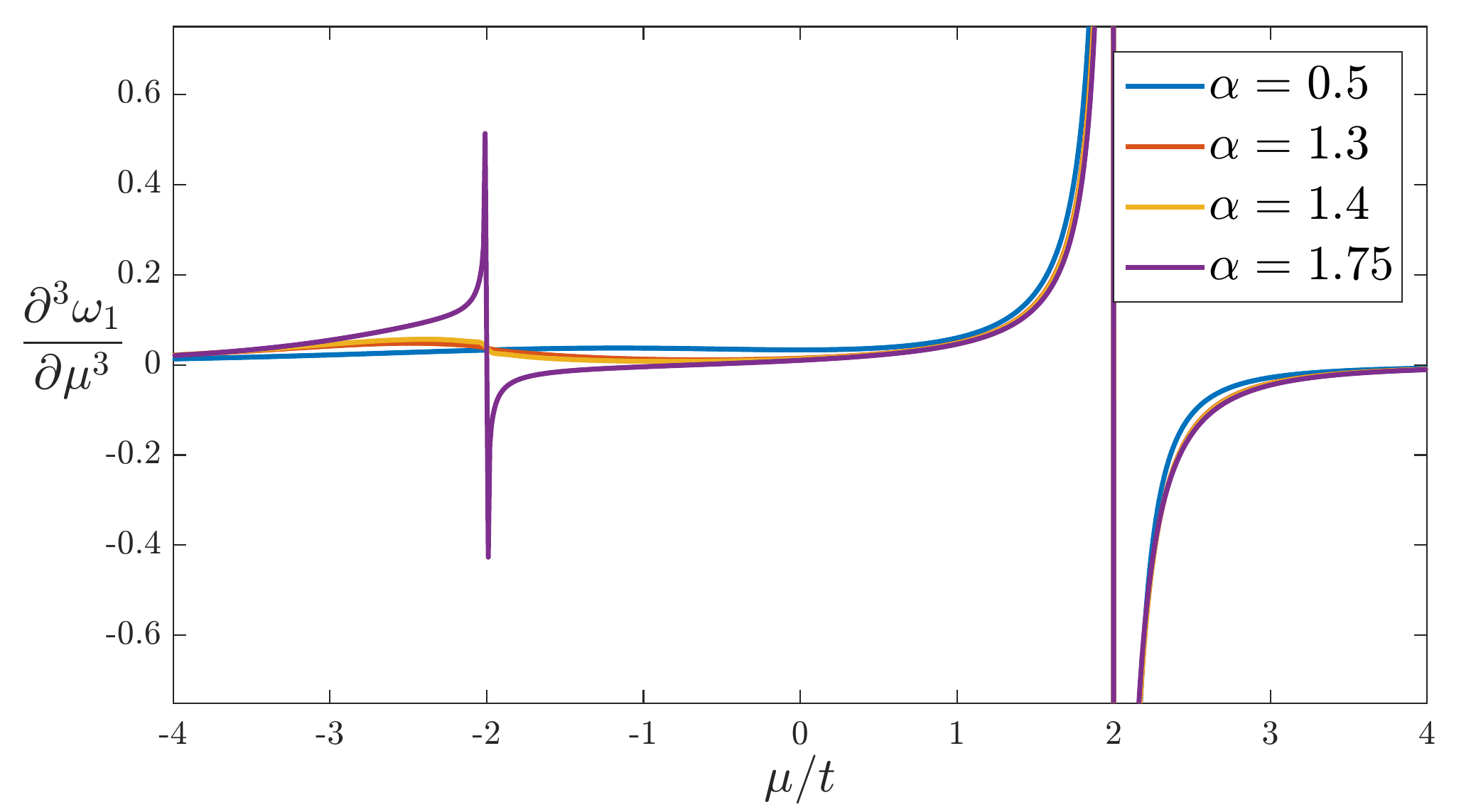} &
\subfigimg[width=1\linewidth]{(d)}{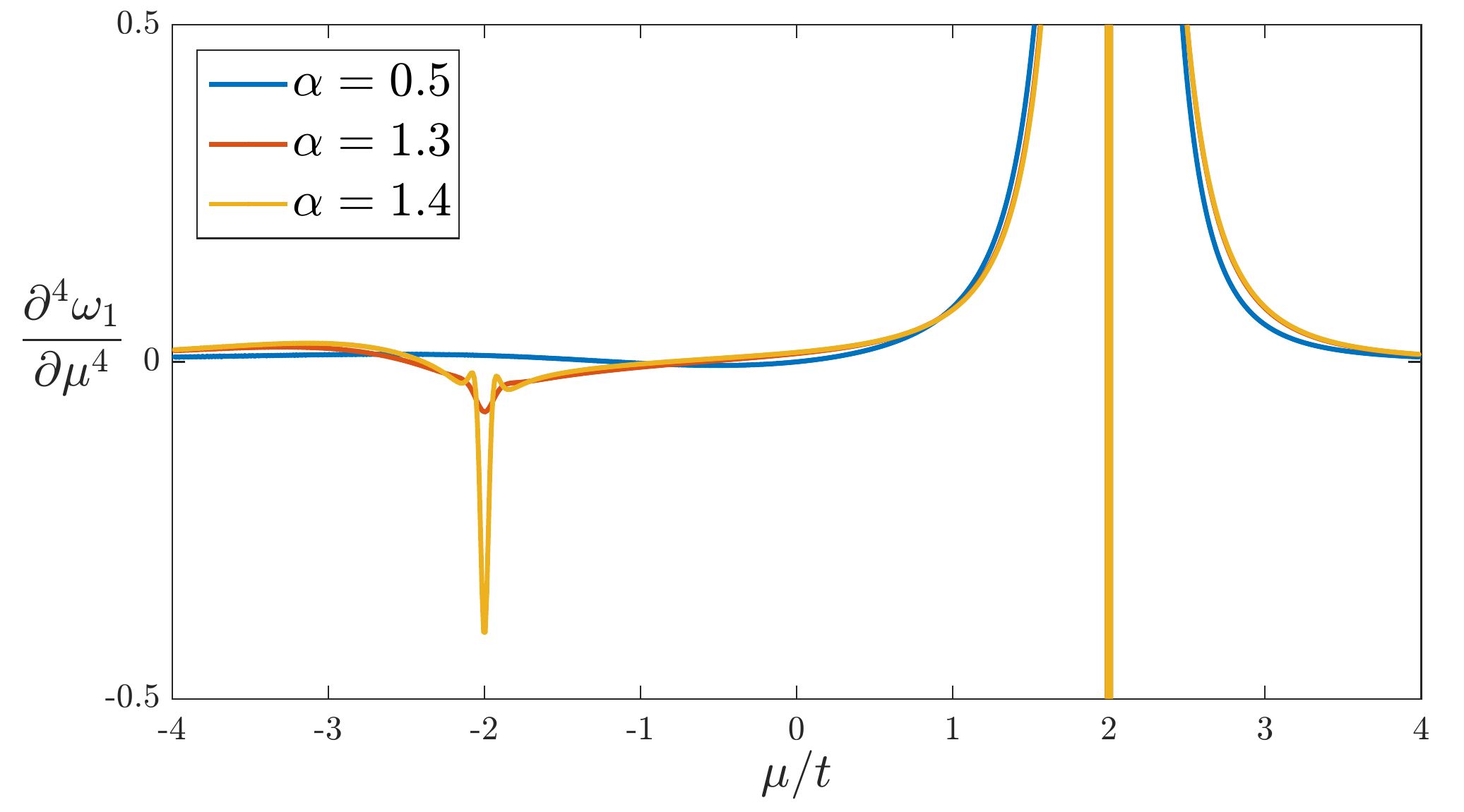}\\
\subfigimg[width=1\linewidth]{(e)}{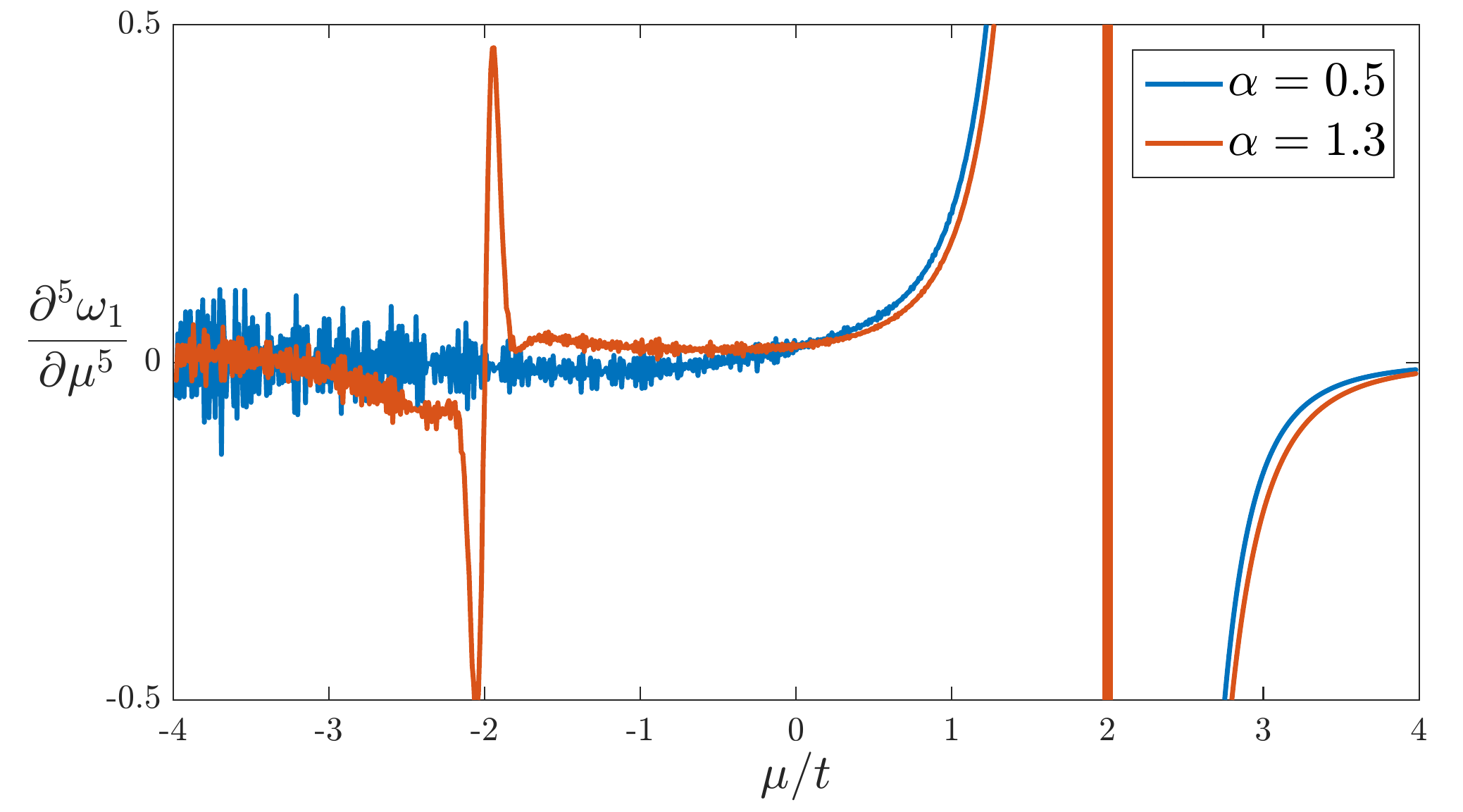} & 
\subfigimg[width=1\linewidth]{(f)}{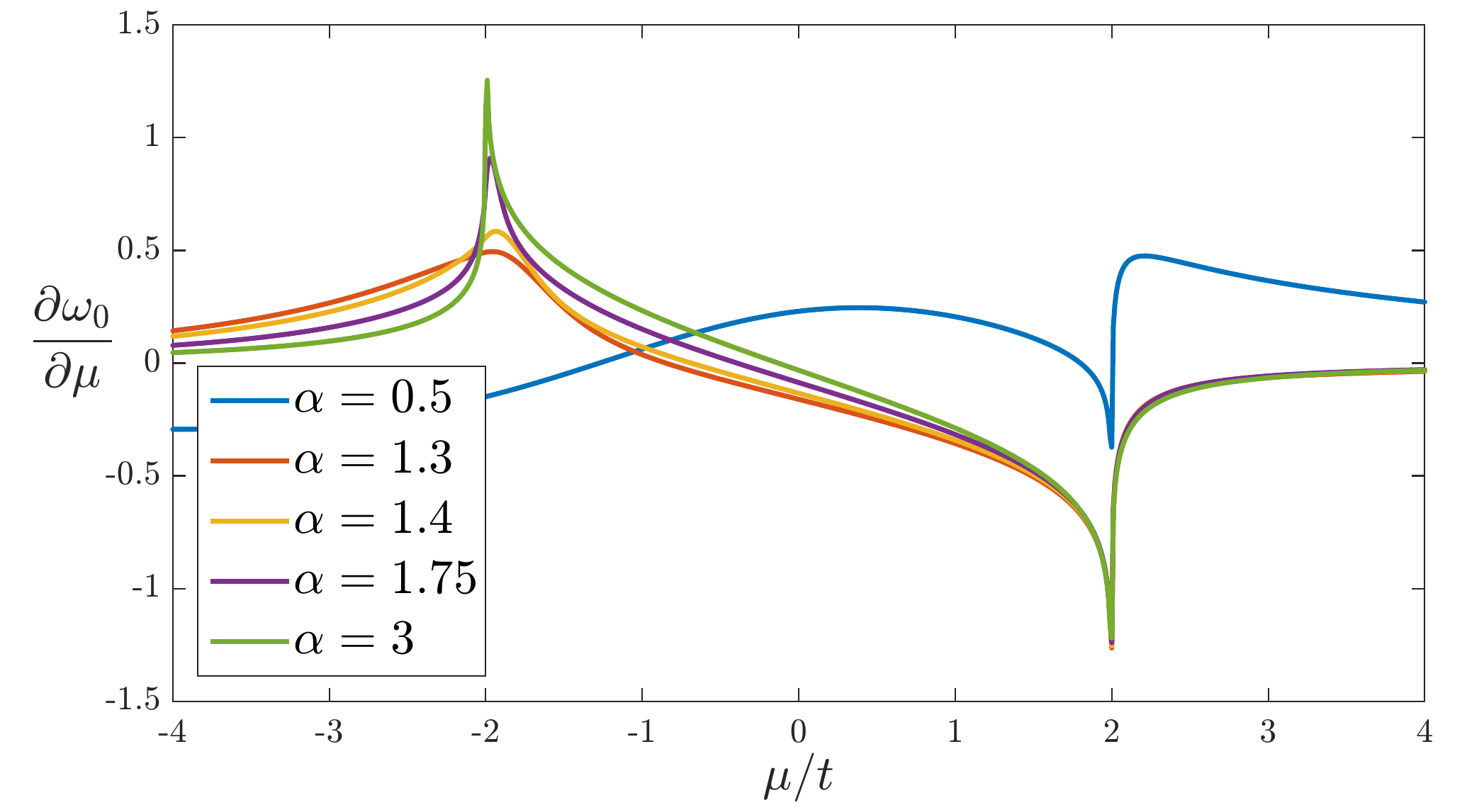}
\end{tabular}
\caption{(a) the bulk contribution $\partial \omega_1/\partial \mu$ and its derivatives (b-e) for several values of $\alpha$. The residual contribution is shown in (f).
There are no peaks in (a), since the phase transition is at least of second order. In (b), sharp peaks at $\mu/t=2$ indicate a second-order phase transition for all $\alpha$, while at $\mu/t=-2$ only a sharp peak for $\alpha=3$ occurs (green). Similarly, upon decreasing $\alpha$, we find a third-order phase transition in (c) for $\alpha=1.75$ (magenta), a fourth-order one in (d) for $\alpha=1.4$ (yellow), a fifth-order one in (e) for $\alpha=1.3$ (orange) and none for $\alpha=0.5$ (blue), as predicted by our analysis.  The noise in (e) is due to numerical differentiation. In the residual contribution $\partial \omega_0/\partial \mu$ (f), there are sharp peaks at $\mu/t=2$ for all $\alpha$, while at $\mu/t=-2$ only those for $\alpha=1.75$ and $\alpha=3$ are sharp. \label{Fig:LR_Kit_supp_thermodynamics}}
\end{figure*}

To better capture the behaviour of the bulk and residual contributions $\partial \omega_1/\partial \mu$ and $\partial \omega_0/\partial \mu$, respectively, at $\mu/t=-2$, we plot here those terms as function of $\mu$ by taking cuts at several values of $\alpha$. In Fig. \ref{Fig:LR_Kit_supp_thermodynamics}, we plotted the bulk contribution as well as its derivatives for the values $\alpha=0.5 , 1.3 , 1.4 , 1.75, 3$, i.e. at the points $\alpha$ in the middle of the stairs. The sharp peaks indicate a phase transition, and indeed we find those peaks exactly for the derivatives predicted by our analytical calculations, i.e. second order for $\alpha=3$ [Fig. \ref{Fig:LR_Kit_supp_thermodynamics} (b)], third order for $\alpha=1.75$ [Fig. \ref{Fig:LR_Kit_supp_thermodynamics} (c)], fourth order for $\alpha=1.4$ [Fig. \ref{Fig:LR_Kit_supp_thermodynamics} (d)] and fifth order for $\alpha=1.3$ [Fig. \ref{Fig:LR_Kit_supp_thermodynamics} (e)]. Due to numerical limitations, we could not analyse higher-order derivatives [notice the noise in Fig. \ref{Fig:LR_Kit_supp_thermodynamics} (e)]. Notice also that for $\mu/t=2$, the phase transition remains second order for all values of $\alpha$. These phase transitions can in principle be detected by measuring the density of states because they are related to the correlation functions. Then, one can differentiate the density of states as function of $\mu$ to detect these peaks.

The boundary contribution, given in Fig. \ref{Fig:LR_Kit_supp_thermodynamics} (f), reveals a clear phase transition at $\mu/t=2$. However, the phase transition at $\mu/t=-2$ becomes less apparent when decreasing $\alpha$ and disappears when $\alpha<1$. Finite-size effects start to play a more important role when decreasing $\alpha$, as can be clearly observed in the spectra in Fig. \ref{Fig:Supp_spectrum}, where the zero-energy edge states start to hybridize, acquiring a finite energy for $\mu/t>-2$. In addition, the first and higher excited states get separated from the bulk (green and magenta line in the plot for $\alpha=1.3$ in Fig. \ref{Fig:Supp_spectrum}, leading to multiple edge states for $\mu/t<0$. All these finite-size effects will "blur out" the phase transition at $\mu/t=-2$. In the thermodynamic limit, the MZMs are at zero energy in the topological phase $|\mu/t|<2$ and merge into a massive non-local Dirac fermion for $\mu/t<-2$.



\begin{figure*}[htb]
\centering
\includegraphics[width=0.8\textwidth]{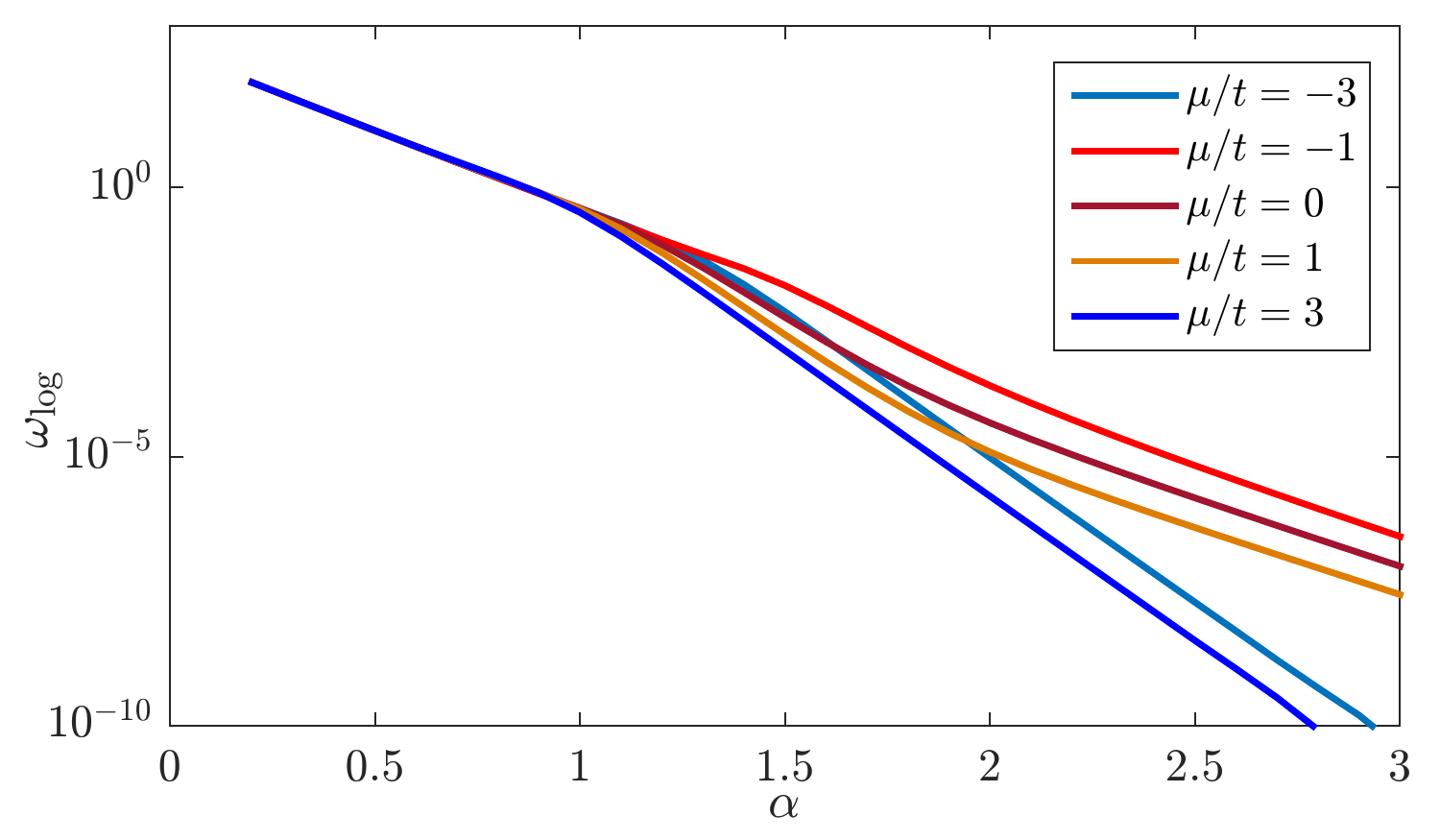}
\caption{The log-term from the expansion of the grand potential as function of $\alpha$ for several values of $\mu$. The reddish lines are for the values of $\mu$ in the topological phase and the blueish lines are for the trivial phase.\label{Fig:LR_Kit_supp_log_term}}
\end{figure*}

The bulk contribution to the grand potential scales linearly with the system size, while the residual contribution contains all the subleading terms, like a $\log(N)$, a constant, a $1/N$, etc. All the terms subleading to the constant one vanish for very large chains. However, the log-term, which is only relevant for small values of $\alpha$ and absent in the SRKC, starts to dominate over the constant term at some point. An accurate separation of the $\log(N)$ from the constant term requires the study of enormously large system sizes (due to the slow growth of the logarithm function), which is numerically not possible to do by brute force. If we still try to separate the constant and log-terms by fitting the grand potential according to $\Omega=N\omega_1+\omega_{\log}\log(N)+\omega_0$ for $N=100$ to $N=2000$, we find that the log-term decays exponentially for non-critical values of $\mu$, as shown in Fig. \ref{Fig:LR_Kit_supp_log_term}, confirming the fact that the log-term is only relevant for small values of $\alpha$. 

\section{Scaling Analysis of End-to-end Correlations}

The phase transition on the boundary can also be captured by considering the end-to-end correlations $\langle c_1^\dagger c_N^\dagger\rangle$ [see Fig.~\ref{Fig:LR_Kit_surf_dw0dmu}] where yellow (blue) indicates strong (small to no) correlations. Here, one sees that for $\alpha>2$, the topological phase with MZMs ($|\mu/t|<2$) has strong end-to-end correlations, because the MZMs at both ends of the chain are combined to form a non-local fermionic excitation. For $\alpha<1$, the correlations are strong everywhere due to the long-range pairing that connects all sites. If $1<\alpha<2$, finite correlations for $|\mu/t|>2$ start to develop, because the correlation functions do not decay fast enough for the system sizes considered, and finite-size effects  become important. Scaling analysis indicates that for all $\alpha>1$, in the thermodynamic limit, the end-to-end correlations are large for $|\mu/t|<2$ and small for $|\mu/t|>2$. This would also imply a first-order phase transition $\mu/t=-2$ for $\alpha$ down to one (as already suggested by the analysis in Fig. 2a in the main text). 

\begin{figure*}[t]
\centering
\includegraphics[width=1.5\columnwidth]{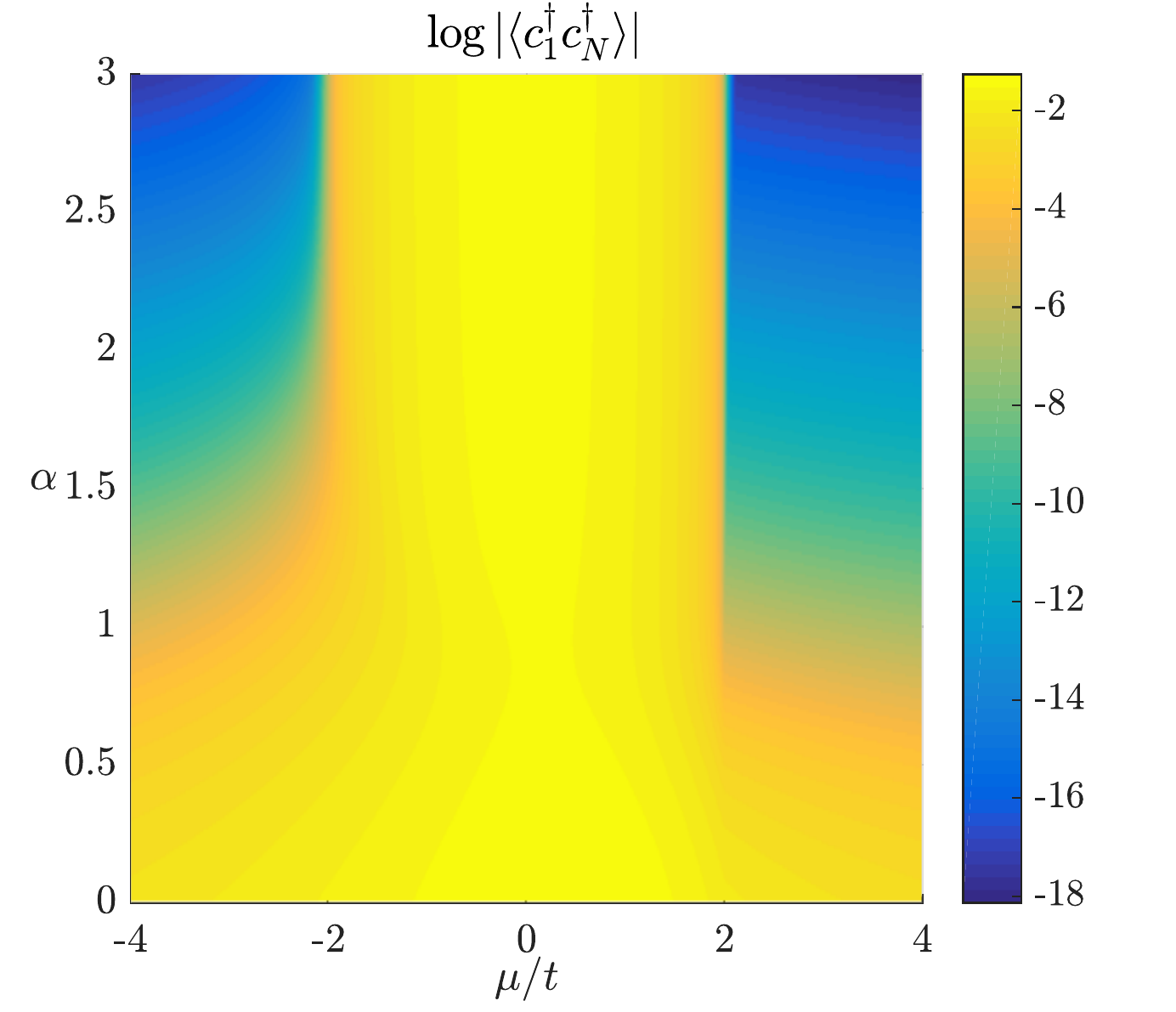}
\caption{End-to-end correlations in the $\mu/t$-$\alpha$ plane. One observes that below $\alpha = 1$, strong correlations appear all over the phase diagram. 
\label{Fig:LR_Kit_surf_dw0dmu}}
\end{figure*}

\begin{figure*}
\centering
\includegraphics[width=1\textwidth]{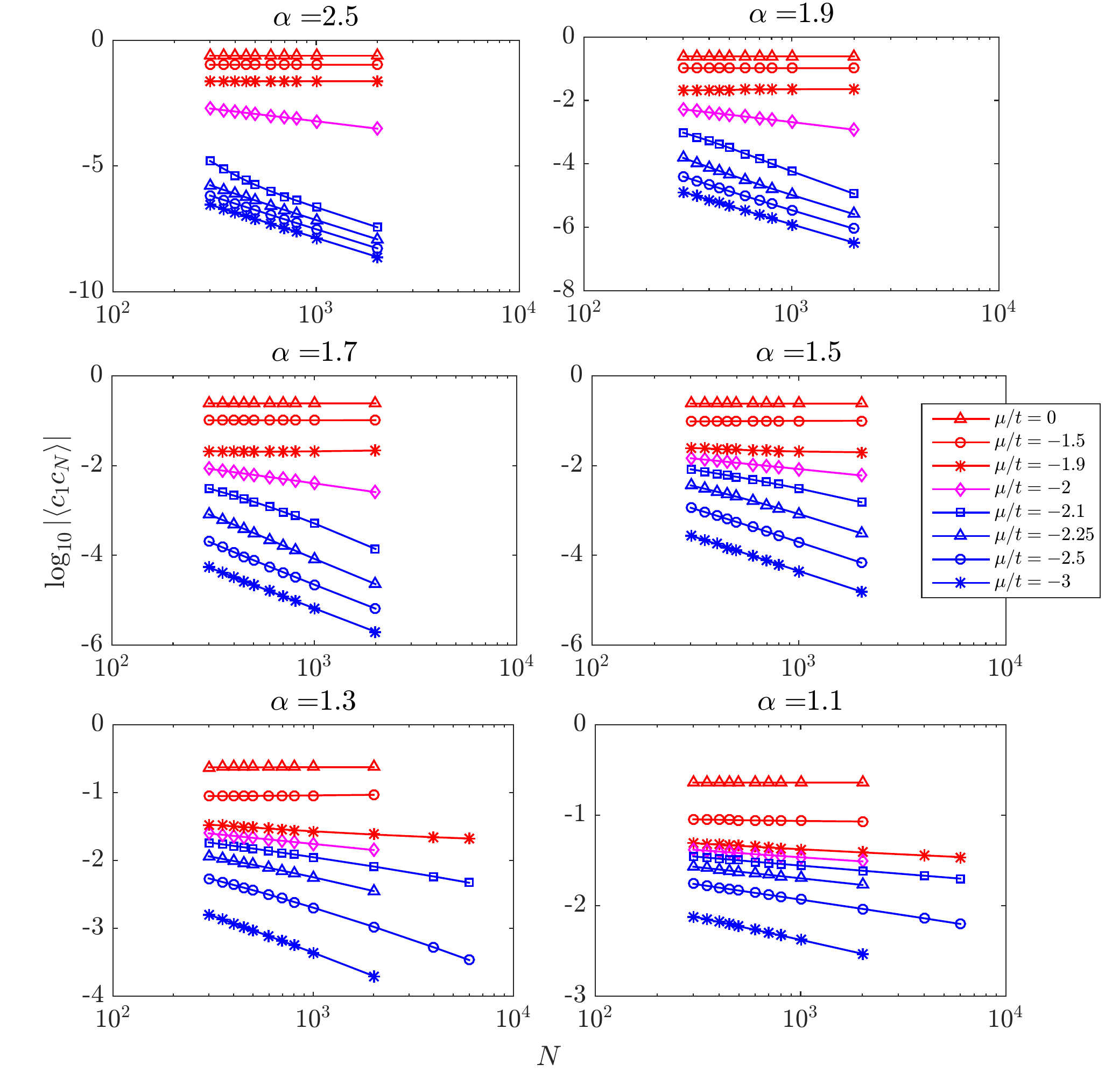}
\caption{The scaling of the end-to-end correlation functions for several values of $\alpha$ and $\mu/t$. The red (blue) lines give the scaling of points inside the topological (trivial) phase. The magenta line corresponds to points on the transition line $\mu/t=-2$. \label{Fig:LR_Kit_supp_N_scaling_c1cN}}
\end{figure*}

The end-to-end correlation functions provide thus a good indication of the existence of non-local fermionic excitations at the edges. Hence, by determining the particular scaling behaviour of these correlation functions we obtain information about the existence of these edge states in the thermodynamic limit. According to our scaling analysis, depicted in Fig. \ref{Fig:LR_Kit_supp_N_scaling_c1cN}, the  end-to-end correlations $\langle c_1 c_N \rangle$ in the trivial phase seem to scale to zero, although very slowly for $\alpha<1.5$, and in the topological phase they seem to converge to some finite value. Some questions can be raised regarding the scaling analysis for $\alpha=1.1$, because the line for $\mu/t=-1.9$ has not converged yet up to $N=6000$. However, it probably will converge and show a similar behaviour as for $\alpha=1.3$ when going to larger system sizes, but we could not access this regime. One would need to go to much larger system sizes to confidently draw conclusions from the scaling analysis.

\end{document}